\documentstyle[epsfig]{article}
\def\cpc#1#2#3{19#3 {\em Comp.\ Phys.\ Commun.}~{\bf#1} #2}
\def\hepex#1#2{19#1 hep-ex/#1#2}
\def\hepph#1#2{19#1 hep-ph/#1#2}

\def\np#1#2#3{19#3 {\em Nucl.\ Phys.}~B {\bf#1} #2}
\def\pl#1#2#3{19#3 {\em Phys.\ Lett.}~{\bf#1B} #2}
\def\pr#1#2#3{19#3 {\em Phys.\ Rev.}~D {\bf#1} #2}
\def\prl#1#2#3{19#3 {\em Phys.\ Rev.\ Lett.}~{\bf#1} #2}
\def\ptp#1#2#3{19#3 {\em Progr.\ Theor.\ Phys.}~{\bf#1} #2}

\def\zp#1#2#3{19#3 {\em Z.\ Phys.}~C {\bf#1} #2}
\def\as{\stub{\alpha}{S}}
\def\beq{\begin{equation}}
\def\BE{Bose-Einstein}
\def\eeq{\end{equation}}
\def\etal{et al.\ }
\def\fref#1{fig.~\ref{#1}}
\def\sref#1{sect.~\ref{#1}}
\def\ssub#1#2{#1_{\mbox{\scriptsize #2}}}
\def\stub#1#2{#1_{\mbox{\tiny #2}}}
\begin{document}
\title{Colour reconnection and Bose-Einstein effects
\footnote{Talk at Phenomenology Workshop on LEP2 Physics, Oxford,
April 1997. Research supported in part by the U.K.\ Particle Physics and
Astronomy Research Council and by the EC Programme
``Training and Mobility of Researchers", Network
``Hadronic Physics with High Energy Electromagnetic Probes",
contract ERB FMRX-CT96-0008.}}

\author{B.R.\ Webber\\
Cavendish Laboratory, University of Cambridge,\\
Cambridge CB3 0HE, UK.}
\maketitle

\begin{abstract}
Final-state interactions and interference phenomena that could affect
the value of the W mass reconstructed from hadronic WW decays at LEP2
are reviewed, and possible areas for future investigation are identified.
\end{abstract}

\section{Introduction}
The accurate measurement of the W boson mass is one of the
primary goals of LEP2. In terms of statistics,
a precision of around $\pm 40$ MeV could eventually be obtained
from reconstructing the mass in fully hadronic WW decays.
However, systematic uncertainties due to interactions and
interference between the W decay products could
degrade this precision substantially \cite{Wmass,evgen}.

The hadronic decay properties of individual, isolated W bosons can
be predicted reliably on the basis of the LEP1 data on Z$^0$
decays \cite{QCDwg}, and these predictions can be tested against
the LEP2 data from semi-leptonic WW decays. The problems in
fully hadronic final states come from the fact that the products
of the two W decays can overlap considerably in space-time \cite{KhoSjo}.
The separation of the W decay vertices at LEP2 energies is
$\sim 0.1$~fm. This is small compared with the
typical hadronic scale of $\sim 1$~fm, so the possibilities for
overlap are great. Final-state interactions and/or identical-particle
symmetrization can then lead to an apparent shift in the W mass.

In this talk, I shall review ideas on the possible
effects of overlap on the W mass measurement, concentrating
on phenomenological developments since the CERN Workshop \cite{Wmass}.
The two main areas of activity are colour reconnection
and \BE\ correlations. I shall discuss these in turn, and
then try to indicate promising directions for future investigation,
both at the current Workshop and beyond.

\section{Perturbative colour reconnection}
\begin{figure}\begin{center}
\epsfig{figure=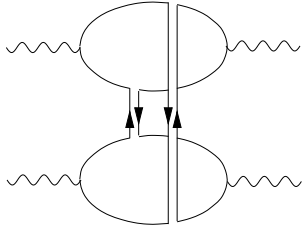, height=3.0cm}
\caption{Lowest-order colour reconnection in WW decay.}
\label{fig_WWgg}
\end{center}\end{figure}

We should start by recalling the nature of colour reconnection
and its expected properties in perturbative QCD.  A relevant
lowest-order contribution to the cross section, $\ssub\sigma{rec}$,
is shown in \fref{fig_WWgg},
where the colour structure is (approximately) displayed by
showing each gluon as a pair of directed colour lines.
The W bosons are colour singlets, but the final states from each
W decay exchange colour. The fact that this graph contains
only two closed colour loops, like the Born graph,
means that it is colour-suppressed, by two powers
of the number of colours $N$,  relative to the leading
${\cal O}(\as^2)$ contributions. Furthermore, the space-time
separation of the decay vertices, of order $1/\stub\Gamma W$,
is large compared with the scale
for hard gluon emission, which is ${\cal O}(1/\stub M W)$.
This leads to a suppression by a further factor of
$\stub\Gamma W/\stub M W$.
Thus the order of magnitude of the perturbative reconnection
rate is \cite{KhoSjo}
\beq
{\ssub\sigma{rec}\over\stub\sigma{WW}}\sim {(\stub C F\,\as)^2\over
N^2}\cdot{\stub\Gamma W\over\stub M W}\sim 10^{-3}\;.
\eeq 
This is small compared with the non-perturbative rate, to be estimated
in the next section. Although this does not automatically mean that
the effect on the W mass is negligible, it is probably within the
uncertainties of the non-perturbative models, and so I shall not
discuss it further here.

\section{Non-perturbative reconnection}
At the hadronization scale of distances ($\sim 1$ fm) the 
space-time separation of the W decay vertices ceases to be so
important, and reconnection can take place wherever the hadronization 
region of the two Ws overlap. This viewpoint shows clearly that a 
space-time picture of the development of the final state becomes essential 
for a realistic treatment of reconnection at the hadronization level. 

\begin{figure}\begin{center}
\epsfig{figure=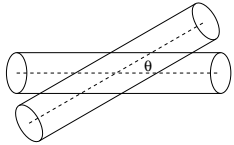, height=3.0cm, angle=270}
\caption{Hadronization region for WW final state.}
\label{fig_WWtubes}
\end{center}\end{figure}

Following hadronic decay of the Ws, there is a perturbative parton 
showering phase of development, in which partons radiated by the initial 
quark and antiquark from each decay spread out and define the regions in 
which hadrons will subsequently form.  These regions are of limited
transverse size but can extend much further in the directions of motion 
of the initial partons, because of time dilation. Typically, therefore,
the final states from the two W decays each have a predominantly
two-jet structure extending some tens of fermis along the directions
of the jet axes but only one or two fermis transverse to it
(\fref{fig_WWtubes}).  Thus there can be a large overlap between the
hadronization products of the two decays, enhancing the probability 
of reconnection.  Notice that the overlap will depend strongly on the 
angle $\theta$ between the decay axes.  We shall discuss this effect 
more quantitatively in \sref{nonsing}.

All the commonly-used models for non-perturbative colour reconnection
are based on a space-time picture in which reconnection is a local
phenomenon.  Objects are formed at the hadronization stage via a local 
interaction which may combine products of the two W decays in regions 
where they overlap.  We may then distinguish two classes of models, 
according to the types of combinations that are permitted.  In {\em 
singlet} models only colour-singlet objects can be formed, whereas in 
{\em non-singlet} models there is no such constraint.

\subsection{Colour singlet models}
According to the above classification, all the reconnection models based
on string hadronization proposed by Khoze and Sj\"ostrand \cite {KhoSjo} 
are singlet models, since each string is a colour singlet. 
Within this framework they investigated two classes of models, called type 
I and type II in analogy to the two types of superconducting vortices which
could correspond to colour strings, assuming the latter are formed by a QCD 
dual Meissner effect.
In the type I scenario the strings/vortices have a significant transverse 
extension, of the order of one fermi, and the probability of reconnection 
depends on the volume of overlap of the strings formed in the two W decays.
In type II, the strings are of negligible thickness, but they reconnect with 
unit probability if they intersect each other at any time between their 
formation and hadronization.

The main alternative to the string hadronization model is the cluster model,
in which quarks and gluons from the parton showers combine locally
into clusters, which are much less extended and less massive objects
than strings, typically light enough to decay more or 
less isotropically into a small number of hadrons each.

The most widely used cluster models have also been colour singlet models,
in which only singlet combinations of partons (in practice, quarks and 
antiquarks) are allowed to form clusters.  In the limit of a large number 
of colours, every quark or antiquark produced in the parton shower
has a unique colour-connected  partner with which it can be clustered,
while every gluon has a colour and 
an anticolour index, each uniquely connected to another parton. In the model 
used in the HERWIG Monte Carlo program \cite{HW51,HW59},
after showering the gluons are split 
non-perturbatively into quark-antiquark pairs, so that each may form a 
colour-singlet cluster with its colour-connected partner.  A strong point
of this scenario is that the perturbative parton shower has the property 
of {\em preconfinement}, which means that the resulting spectrum
of cluster sizes and masses peaks at small values and is universal,
i.e.\ independent of the nature and scale of the 
hard process that initiated the showers \cite{AmaVen,BCM}.

As in the string model, local colour reconnection in the cluster model 
is natural once a space-time picture of the development of the final
state is included.  If, for example, a quark finds that its nearest
neighbouring antiquark of the right colour is not its colour-connected 
partner, then it may well prefer to cluster with the former.  Note that 
this requires us to go beyond the large-$N$ limit, since as $N\to\infty$
it becomes vanishingly improbable that any antiquark other than the 
colour-connected one will have the right colour to form a singlet.  In 
the real world of $N=3$ we would expect the probability to be $1/9$, 
since even if the antiquark has the right colour, two out of the three
same-colour combinations are octets rather than singlets.

\begin{figure}\begin{center}
\epsfig{figure=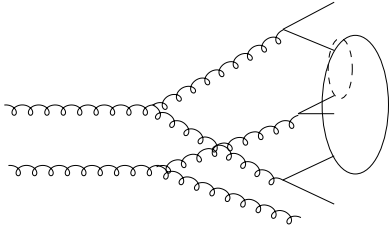, height=3.0cm}
\caption{HERWIG model for colour reconnection. Solid: usual clustering.
Dashed: possible alternative clustering.}
\label{fig_reconnect}
\end{center}\end{figure}

A scenario of this type has been implemented in HERWIG,
version 5.9 \cite{HW59}.  
After parton showering and gluon splitting, at the start of the cluster 
formation phase, the program looks for switches of colour connections
between clusters that would reduce the space-time extension of the clusters,
as illustrated in \fref{fig_reconnect}.  If one is found, reconnection
occurs with probability set by a parameter {\tt PRECO}, default value $1/9$.

The implications of colour-singlet string and cluster models for the
W mass determination were investigated during the CERN Workshop \cite{Wmass}. 
The effects of colour reconnection were generally found to be small.
Compared with the equivalent model without reconnection between the two Ws,
obtained for example by enhancing the W lifetime to prevent any space-time
overlap between their decays, the mass shifts obtained were of the order 
of 100 MeV times the WW reconnection rate per event (typically 10-50\%). 
These results are, however, sensitive to the way in which the W mass is
deduced from the data, and it would certainly be of interest to repeat the
model studies using the actual methods now being employed in the LEP
experiments.

Note that in all the above models there is a possibility of reconnections
amongst the hadronization products of a single parton shower, not only
between those from different boson decays. This is perhaps most natural
in the cluster model, but in the string model a single decay can also give
rise to multiple string segments or to a single string that intersects
itself.  Thus a retuning of the model parameters to agree with LEP1 data
on single Z boson decays is really required when reconnection is introduced.
This has not been done in detail, but there is no reason to suppose that
it would enhance the mass shifts due to reconnection in WW events.

\subsection{Non-singlet models}\label{nonsing}
Although the formation of colour-singlet strings or clusters may
appear the most plausible first stage of hadron production, there are
good reasons to believe that this is not the whole story. One reason
is that partons can still end up far away from their colour-connected
partners after parton showering.  In this case they may prefer to
interact non-perturbatively with nearby partons of the wrong colour
and form hadrons more indirectly.

Experimental evidence for an important non-singlet component of
hadronization comes from the failure of singlet models to account for
the data on production of heavy quarkonia.
The scale for the formation of the heavy quarks themselves is set by
the quark mass, whereas that for formation of the observed quarkonium
states is set by the binding energy of the latter.  There is thus a
particularly large gap in this case between the scale at which parton
showering effectively stops and that at which the final hadrons are
formed.  The hypothesis that the heavy quarks must be in a singlet
state at the end of showering in order to bind greatly
underestimates the amount of quarkonium production
in hadron collisions \cite{CDF,E789}. A large, indeed dominant,
colour-octet component is also required \cite{Beneke}. There
are also indications of an octet contribution to $J/\psi$
production in Z$^0$ decays \cite{OPALJ,L3J}.
The octet states must then evolve by non-perturbative gluon emission into
the observed singlets.  In the case of quarkonium there is plenty of time
for this to happen, owing to the scale discrepancy mentioned above.
In the case of light-quark hadrons there is not a mismatch of scales,
but the hadronization timescale is just as long, and so it would be
surprising if a similar mechanism did not contribute at some level.

The only general hadronization model available at present which
includes a non-singlet component is that of Ellis and Geiger \cite{EllGei},
which is based on a transport-theoretical treatment of
parton showering and cluster formation.  A novel feature of the model
is that it uses an effective Lagrangian containing both partonic and
hadronic degrees of freedom to generate the parton shower.  The two
components have scale-dependent couplings that imply dominance of
the partonic degrees of freedom at short distances  and of the hadronic
ones at long distances. As a consequence, whenever partons start
to move too far away from their neighbours, cluster formation begins
and prevents them from becoming widely separated, as required for
colour confinement.

In \cite{EllGei}, three scenarios for the mechanism of cluster formation
are considered.  The first two are, roughly speaking, colour-singlet models
similar to that used in HERWIG, discussed above.  Like the other singlet
models, they imply relatively small shifts in the W mass due to colour
reconnection.  However, the third scenario, which Ellis and Geiger
call ``colour-full", includes non-singlet clustering:
nearest-neighbour parton pairs in any colour combination are clustered
if their separation exceeds a critical value.
The net colour of the cluster, if any, is carried off by a secondary
parton, and the process continues until all partons have been clustered.
This scenario leads to much larger mass shifts, possibly as large as
400 MeV (in the positive direction).

In view of the big difference between the W mass shifts in the
colour-singlet models and the only available non-singlet model, and
the experimental evidence for a non-singlet mechanism in quarkonium
production, it is clearly important to investigate the predictions
of the Ellis-Geiger colour-full scenario as fully as possible.
First of all, as with the singlet models in which colour reconnection
has been introduced as a new feature, one should tune the parameters of
the model to achieve optimal agreement with LEP1 data.  A point of
particular interest would be to compare the predictions of
this model\footnote{Predictions for quarkonium hadroproduction
have become available since the Workshop \cite{Gei97}.} and the
singlet models with data on quarkonium production in both
hadron-hadron and electron-positron collisions.

\begin{figure}\begin{center}
\epsfig{figure=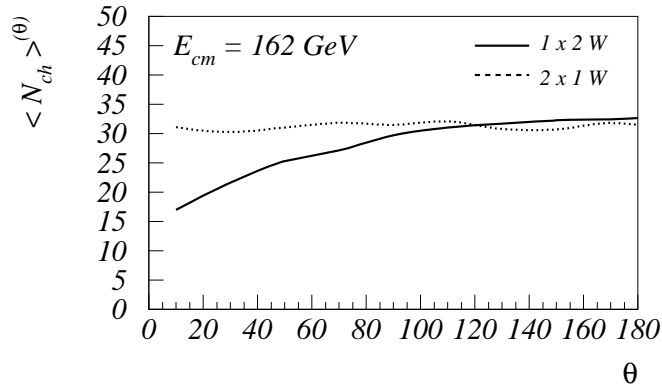, height=5.0cm}
\caption{Mean charged multiplicity as a function of the angle
$\theta$ defined in \fref{fig_WWtubes}.}
\label{fig_EGmult}
\end{center}\end{figure}

\begin{figure}\begin{center}
\epsfig{figure=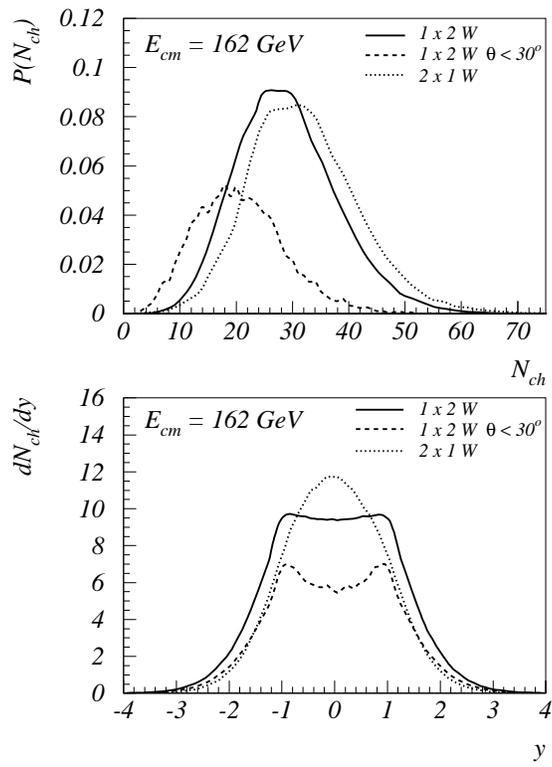, height=10.0cm}
\caption{Charged multiplicity and rapidity distributions.}
\label{fig_EGnydist}
\end{center}\end{figure}

In addition, the Ellis-Geiger model predicts some striking reconnection
effects in WW events, besides the large W mass shift \cite{EllGei}. 
In particular, a substantial reduction in the hadron multiplicity coming
from the overlap of the hadronization regions of the two Ws is expected.
Since the overlap increases as the angle $\theta$ between the jet
axes of the two W decays decreases (see \fref{fig_WWtubes}), this
implies a reduced multiplicity at small values of $\theta$, as shown
\fref{fig_EGmult}. The decrease is predicted to be mainly in
the region of central rapidities relative to the thrust axis
(\fref{fig_EGnydist}). 


It is important to establish how well such an effect could be seen in
the data, taking into account the fact that events with small values of
$\theta$ tend to be removed by cuts designed to eliminate QCD background.
Presumably the multiplicity in the overlap region falls because the
enhanced parton density there leads to the production of lower-mass
clusters, so that the hadron density does not rise proportionately.
In non-singlet models, secondary parton emission from non-singlet
clusters contributes to increasing the parton density, enhancing
the effect.  In the Ellis-Geiger model, it is noticeable
that in single Z$^0$ decay the colour-full scenario has the highest
parton and hadron densities in the central rapidity region, where
most overlap will occur in WW events.  Here again, however, it will
be important to achieve good agreement with the LEP1 data before
quantitative predictions for WW can be made.

A reduction of multiplicity at low rapidities in overlapping W decays
would be expected to increase the measured value of the W mass in WW
events, as observed in \cite{EllGei}. This is because the misassignment
of wide-angle particles amongst jets, which tends to decrease the mass,
will be less important if there are fewer such particles.

In colour-singlet models one would also expect a $\theta$ dependence
of the multiplicity, although presumably much smaller and not necessarily
of the same sign.  Thus a thorough investigation of these effects would
be worthwhile as a direct test of the models in WW events.  In the
HERWIG model discussed above, the amount of colour reconnection can be
varied by changing the input parameter {\tt PRECO}; setting this to
unity would correspond to ignoring colour altogether during cluster
formation. This goes some way towards a non-singlet mechanism,
although without the secondary parton emission of the Ellis-Geiger
colour-full scenario. If an effect strongly correlated with a
W mass shift were seen in a variety of models, its measured value
could serve to calibrate the magnitude of the mass shift.

\section{Modelling \BE\ effects}
\BE\ correlations between identical bosons (in practice, pions)
are also a potential source of a W mass shift in WW events. There
are almost four times as many identical-pion pairs in the WW final
state as in a single W decay, and so some kind of non-linear effect,
possibly leading to a mass shift, would be expected.  The problem with
estimating such effects is that they are purely quantum mechanical in
nature, whereas the Monte Carlo programs used to generate simulated
events are based on classical models.

In fact it is not quite true that classical Monte Carlo programs
cannot simulate purely quantum phenomena.  Spin correlations that
embody the Einstein-Podolsky-Rosen ``paradox" can be simulated in full
because the program does not have to respect causality: information
can be propagated backwards in time as long as the program is not
performing a real-time simulation.  In this way acausal correlations
can be incorporated \cite{Kno}. 
However, the problem with simulating \BE\ effects is not so
much with causality as with the computational complexity
of symmetrizing with respect to a very large number of variables.
In response to this problem, several methods for grafting
\BE\ correlations onto existing event generators have been
proposed. We deal briefly with each of them in turn below.

\subsection{Redistribution method}
The most developed technique is that proposed and incorporated into
the JETSET generator by Sj\"ostrand \cite{JS74,LonSjo}.  In this
approach the momenta of identical final-state particles are redistributed
(shifted) to reproduce the expected two-boson momentum correlations.
The advantage of this method is that, since it involves no reweighting
of events, it remains a true (unit-weight) event generator and does
not suffer any loss of efficiency.  The main disadvantage is that the
momentum shifts spoil overall energy-momentum conservation and so one
has to modify also some momenta of non-identical particles
in order to compensate for this.
In addition,  \BE\ correlations involving more than
two particles are omitted. Nevertheless the approach is a convenient way
of incorporating the two-particle correlations observed at LEP1 and
investigating their possible consequences in WW events.  It was found
\cite{LonSjo} that the implications for the W mass measurement could
be quite severe.

A worrying feature of the current approaches to \BE\ effects -- the
redistribution method and the reweighting methods
to be discussed below -- is that they take little account of the space-time
development of the final state.  Identical bosons are supposed to originate
from a common production region characterized by a few parameters,
which cannot do justice to the complexity of the actual hadronization
region, especially in WW final states. In reality, the effects of overlapping
regions discussed above for colour reconnection will also be relevant for
\BE\ effects.  Identical pions from different W decays would only
be expected to interfere if there is significant overlap. The components
of the two-pion correlation function due to same-W and different-W
interference can be separated by comparing fully hadronic
and semi-leptonic final states.  At present there
is no evidence of a different-W \BE\ effect in the
data \cite{Delphi}.  Here again it would be most interesting to
explore the correlation between such an effect
and the W mass shift in a variety of models. 

\subsection{Reweighting methods}
Another method for imposing \BE\ correlations on classically
simulated events is by weighting each event with
a symmetrized weight function.  In principle, all the multiparticle
correlations could be fully included in this way.  The problem is that,
even if we knew how to compute the correct weight function, its calculation
would be too laborious, involving a sum over all permutations of
particles. In addition, the distribution of event weights would probably
be so broad that the simulation would become hopelessly inefficient.
This has led to the investigation of `partial symmetrization' procedures
that aim to include the most important permutations for each event.
Two such procedures are discussed in more detail below.

\subsubsection{Clustering.}
One possible way of limiting the amount of computation is to
organize the identical particles into {\em clusters} such that
the significant weight factors are likely to be limited to
permutations within clusters. The invariant measure of closeness
in phase space, for two identical bosons $i$ and $j$ of
mass $m$, is
\beq
Q_{ij}^2 = -(p_i-p_j)^2 = M_{ij}^2-4m^2\;.
\eeq
The \BE\ weight will approach unity for large $Q_{ij}$,
and so one defines a cluster as a set of identical pions such that
each member $i$ has at least one neighbour $j$ with $Q_{ij}<Q_0$.
Here $Q_0$ is a parameter, ideally such that $RQ_0\gg 1$ where
$R$ is the size of the source region. Then only permutations within
clusters are considered in computing the \BE\ weights.

This approach has been applied in \cite{JadZal}. A rather
small value of the cluster cutoff, $Q_0=0.2$ GeV, was used,
thereby limiting the typical cluster size and simplifying the
weight calculation. It would be good to investigate the stability
of the results with respect to variation of $Q_0$.
The cluster weights were computed using
the model of \cite{Biya}. To keep the total cross section
and the mean value of the pion multiplicity $n$ unchanged, a
further overall weight factor of $c\lambda^n$ was applied, with
$c$ and $\lambda$ determined retrospectively.  The conclusion of
\cite{JadZal} is that the shift in the reconstructed W mass
due to \BE\ correlations is not more than 30 MeV.

\subsubsection{Limited permutation.}
Another possibility is to organize permutations into those of
exactly 2, 3, 4,\ldots\ identical particles, and then take into
account only those up to some maximum number $K$ \cite{FiaWit}.
The validity of the truncation can be checked to some extent,
by comparing the results for $K-1$ and $K$. In \cite{FiaWit},
this method has been applied to minimum-bias proton-antiproton
collisions; results are presented for $K=4$ and 5, showing good
stability. Once again the weights were rescaled by $c\lambda^n$
to keep the total cross section and mean pion multiplicity unchanged.
Clearly it would be interesting to apply this method to WW final states.

\subsection{Other methods}
An important point to bear in mind is that much work on
\BE\ correlations has also been undertaken in the context of nuclear
physics, and we may be able to borrow some ideas from that field.
For example, in \cite{ZWSH} it is proposed to treat the
classical space-time points of origin of pions in Monte Carlo
simulations of heavy ion collisions as sources of Gaussian wave
packets. The two-particle correlation function can then be
computed, with the width of the packets as a free parameter.
Such an approach looks feasible for WW events, and it might
meet the objection that existing models do not take adequate
account of the space-time structure of the process.

Another interesting recent development is a treatment of
\BE\ correlations which emerges naturally from the Lund string model
of hadronization \cite{AndRin}. Here again the space-time structure
is included and therefore a more satisfactory treatment of WW final
states may be possible.

\section{Summary}
Unfortunately, the possible effects of colour reconnection and
\BE\ correlations on the reconstructed W mass are still uncertain.
The necessity for a large colour-octet contribution to quarkonium
production suggests that one should take seriously the possibility
of a non-singlet hadronization mechanism for light
quarks as well, which could lead to a larger mass shift than is
currently estimated using colour-singlet models. As for
\BE\ correlations, studies done so far using the reweighting
method suggest smaller mass shifts than those found with the
redistribution method, although further work is needed to confirm
this\footnote{A comparative study of reweighting schemes
has been completed recently \cite{KKM}; some results
are presented later in these Proceedings \cite{WmWG}.}. 

Some topics on which further work would appear worthwhile are:
\begin{itemize}
\item To update the CERN Workshop studies of colour reconnection
effects in various models, using the methods now being employed
to extract the value of the W mass.
\item To tune the parameters of models which allow reconnection
in single Z$^0$ decays (especially the Ellis-Geiger model), to
optimize agreement with LEP1 data.
\item To add a non-singlet hadronization component to singlet
models, and to investigate how it affects the W mass determination.
\item To study quarkonium production in singlet hadronization
models, to see whether a non-singlet contribution is essential.
\item To establish whether a reduction in multiplicity at small values
of the WW relative decay angle $\theta$ could be seen in the data, after
cuts to remove QCD background.
\item To study the same effect in a variety of models and see
whether any drop in multiplicity at small $\theta$ is correlated
with a W mass shift.
\item To look for a correlation between a different-W \BE\
effect and a W mass shift in various models.
\item To apply the limited-permutation approximation for \BE\
weights to WW final states.
\item To compute \BE\ correlations in WW final states by using
the space-time configurations generated by existing Monte Carlo programs
as sources of Gaussian wave packets.
\end{itemize}
Perhaps the most useful outcome would be to establish a convincing
correlation between some measurable effect and the W mass shift,
which would help us to estimate the latter in a slightly less
model-dependent way. For this purpose the multiplicity at small
$\theta$ and the different-W \BE\ effect look promising.

\goodbreak


\begin{thebibliography}{99}
\bibitem{Wmass}
Kunszt Z \etal (W mass working group) 1996 {\em Physics at LEP2}, CERN 96-01
vol 1 p 141, ed G Altarelli, T Sj\"ostrand and F Zwirner
\bibitem{evgen}
Knowles I G \etal (QCD event generators working group) 1996
{\em Physics at LEP2}, CERN 96-01 vol 2 p 103,
ed G Altarelli, T Sj\"ostrand and F Zwirner
\bibitem{QCDwg}
Nason P \etal (QCD working group) 1996 {\em Physics at LEP2}, CERN 96-01
vol 1 p 249, ed G Altarelli, T Sj\"ostrand and F Zwirner
\bibitem{KhoSjo}
Sj\"ostrand T and Khoze V A \prl{72}{28}{94}; \zp{62}{281}{94}
\bibitem{HW51}
Marchesini G \etal \cpc{67}{465}{92}
\bibitem{HW59}
Marchesini G \etal \hepph{96}{07393}
\bibitem{AmaVen}
Amati D and Veneziano G \pl{83}{87}{79}
\bibitem{BCM}
Bassetto A,  Ciafaloni M and Marchesini G \pl{83}{207}{79}
\bibitem{CDF}
Bailey M W (CDF collaboration) \hepex{96}{08014}, preprint
Fermilab-Conf-96/235-E
\bibitem{E789}
Kaplan D M (E789 collaboration) \hepex{96}{10003}, preprint IIT-HEP-96/4
\bibitem{Beneke}
Beneke M \hepph{96}{05462}, preprint SLAC-PUB-7173
\bibitem{OPALJ}
Alexander G \etal (OPAL collaboration) \pl{384}{343}{96}
\bibitem{L3J}
Acciarri M \etal (L3 collaboration) 1997 preprint CERN-PPE/97-044
\bibitem{EllGei}
Ellis J and Geiger K \pr{54}{1967}{96};
\hepph{97}{03348}, preprint CERN-TH/97-46
\bibitem{Gei97}
Geiger K \hepph{97}{06425}
\bibitem{Kno}
Knowles I G \np{310}{571}{88}; \cpc{58}{271}{90}
\bibitem{JS74} 
Sj\"ostrand T \cpc{82}{74}{94}
\bibitem{LonSjo} 
Lonnblad L and Sj\"ostrand T \pl{351}{293}{95}
\bibitem{Delphi}
Abreu P \etal (DELPHI collaboration) 1997 preprint CERN-PPE/97-30
\bibitem{JadZal}
Jadach S and Zalewski K 1997 preprint CERN-TH/97-29
\bibitem{Biya}
Biyajima M \etal \ptp{84}{931}{90}
\bibitem{FiaWit}
Fia\l kowski K and Wit R  \zp{74}{145}{97}; \hepph{97}{03227}
\bibitem{ZWSH}
Zhang Q H, Wiedemann U A, Slotta C and Heinz U 1997 nucl-th/9704041.
\bibitem{AndRin}
Andersson B and Ringn\'er M \hepph{97}{04383}, preprint LU TP 97-07
\bibitem{KKM}
Kartvelishvili V, Kvatadze R and M\o ller R \hepph{97}{04424},
preprint MC-TH-97-04 rev.
\bibitem{WmWG}
{\em Report of W mass working group}
\end{thebibliography}
\end{document}